\documentclass[12pt]{article}
\usepackage[top=20truemm,bottom=20truemm,left=23truemm,right=23truemm]{geometry}
\usepackage{amsfonts,amssymb,bm,amsmath,cite,graphics,graphicx}
\usepackage{array, booktabs}
\usepackage{slashed}

\newcommand {\del}{\partial}
\newcommand {\tr}{\mbox{tr}}
\newcommand {\Tr}{\mbox{Tr}}

\newcommand {\comm}[2]{\boxbc{#1, #2}}

\newcommand {\pdv}[2]{\frac{\del #1}{\del #2}}

\newcommand {\paren}[1]{\left( #1 \right)}

\newcommand {\boxbc}[1]{\left[ #1 \right]}

\numberwithin{equation}{section}

\allowdisplaybreaks

\begin{document}
\setlength{\oddsidemargin}{0cm}
\setlength{\baselineskip}{7mm}

\begin{titlepage}
\renewcommand{\thefootnote}{\fnsymbol{footnote}}
\begin{normalsize}
\begin{flushright}
\begin{tabular}{r}
KEK-TH-2169
\end{tabular}
\end{flushright}
  \end{normalsize}


\vspace{0.5cm}
    \begin{Large}
       \begin{center}
        {The emergence of expanding space--time and intersecting D-branes\\ 
        from classical solutions in the Lorentzian type IIB matrix model}
       \end{center}
    \end{Large}
\vspace{0.5cm}

\begin{center}
           Kohta H{\sc atakeyama}$^{1,2)}$\footnote
            {
e-mail address : 
hatakeyama.kohta.15@shizuoka.ac.jp},
		   Akira M{\sc atsumoto}$^{3)}$\footnote
		              {
		  e-mail address : 
		  akiram@post.kek.jp},
		  Jun N{\sc ishimura}$^{3,4)}$\footnote
		              {
		  e-mail address : 
		  jnishi@post.kek.jp}, \\
           Asato T{\sc suchiya}$^{1,2)}$\footnote
            {
e-mail address : 
tsuchiya.asato@shizuoka.ac.jp}
           {\sc and}
           Atis Y{\sc osprakob}$^{3)}$\footnote
           {
e-mail address : ayosp@post.kek.jp}\\
      \vspace{0.5cm}

 $^{1)}$ {\it Department of Physics, Shizuoka University}\\
                {\it 836 Ohya, Suruga-ku, Shizuoka 422-8529, Japan}\\
         \vspace{0.15cm}     
 $^{2)}$ {\it Graduate School of Science and Technology, Shizuoka University}\\
               {\it 3-5-1 Johoku, Naka-ku, Hamamatsu 432-8011, Japan}\\
          \vspace{0.15cm}   
 $^{3)}$ {\it Department of Particle and Nuclear Physics, School of High Energy Accelerator Science,}\\
 {\it Graduate University for Advanced Studies (SOKENDAI)}\\
                 {\it 1-1 Oho, Tsukuba, Ibaraki 305-0801, Japan}\\
           \vspace{0.15cm}   
  $^{4)}$ {\it KEK Theory Center,
  Institute of Particle and Nuclear Studies,}\\
  {\it High Energy Accelerator Research Organization}\\
                 {\it 1-1 Oho, Tsukuba, Ibaraki 305-0801, Japan}

\end{center}

\hspace{1.5cm}

\begin{abstract}
\noindent
The type IIB matrix model is a promising candidate for
a nonperturbative formulation of superstring theory.
As such, it is expected to explain
the origin of space--time and matter at the same time.
This has been partially demonstrated
by the previous Monte Carlo studies
on the Lorentzian version of the model, which suggested the emergence
of (3+1)-dimensional expanding space--time.
Here we investigate the same model
by solving numerically the classical equation of motion, 
which is expected to be valid at late times since the action 
becomes large due to the expansion of space.
Many solutions are obtained by the gradient descent method
starting from random matrix configurations,
assuming a quasi-direct-product structure
for the (3+1)-dimensions and the extra 6 dimensions.
We find that these solutions generally admit 
the emergence of expanding space--time and 
a block-diagonal structure in the extra dimensions,
the latter being important for the emergence of intersecting D-branes.
For solutions corresponding to D-branes
with appropriate dimensionality,
the Dirac operator
is shown to acquire a zero mode in the limit of infinite matrix size.
\end{abstract}
\vfill
\end{titlepage}
\vfil\eject

\setcounter{footnote}{0}


\section{Introduction}
\setcounter{equation}{0}
Superstring theory has been investigated intensively as a unified theory 
including quantum gravity.
However, in its perturbative formulation that has been established so far,
there exist tremendously many stable vacua 
corresponding to
various space--time dimensionality, gauge groups and matter contents. 
Also, it is known that 
the cosmic singularity at the beginning of the Universe cannot
be resolved within perturbative superstring theory
\cite{Lawrence:2002aj,Liu:2002kb,Horowitz:2002mw,Berkooz:2002je}.
In order to address these issues, nonperturbative formulation 
is clearly needed, 
and the matrix models \cite{Banks:1996vh,Ishibashi:1996xs,Dijkgraaf:1997vv} 
have been proposed as its candidates.

In this paper, we focus on the type IIB matrix model \cite{Ishibashi:1996xs},
which is distinctive in that 
not only space but also time
emerges dynamically from the matrix degrees of freedom.
Indeed, it was shown by Monte Carlo simulation
that (3+1)-dimensional expanding space--time appears 
from the Lorentzian version of the 
model \cite{Kim:2011cr}.
The time scale that has been probed by the simulation is 
typically of the order of the Planck scale.
This line of research has been continued in 
Refs.~\cite{Ito:2013ywa,Ito:2015mxa,Ito:2017rcr,Azuma:2017dcb,Aoki:2019tby,Nishimura:2019qal}, 
whereas the Euclidean version
has been investigated
in Refs.~\cite{Anagnostopoulos:2013xga,Anagnostopoulos:2017gos}
and references therein.

In order to probe the space--time at later times than the Planckian time, 
we need to increase the matrix size by many orders of magnitude, 
which is not feasible in numerical simulations.
On the other hand, it is expected that the classical approximation 
is valid at late times
because the action becomes large due to the expansion of space.
Various classical solutions that allow
cosmological interpretations have been 
found \cite{Kim:2011ts,Kim:2012mw,Chaney:2015ktw,Steinacker:2017vqw},
and a systematic method for constructing classical solutions 
was also developed \cite{Kim:2012mw}.
(See also 
Refs.~\cite{Steinacker:2019awe,Steinacker:2019fcb,Manolakos:2019tdf} for 
related papers in this direction.)
However, the class of solutions that has been constructed so far
looks too simple to describe our real world.
%

In fact, solving the classical equation of motion
in the type IIB matrix model
is quite different from ordinary classical dynamics,
in which one solves
a differential equation with some initial conditions.
This is because 
the notion of time does not exist from the outset
in this model, and it should emerge dynamically
from the matrix degrees of freedom.
Here we propose a method 
based on the gradient descent method,
which enables us to generate infinitely 
many classical solutions starting from random
matrix configurations.
In particular, we assume 
a ``quasi'' direct-product 
structure for the (3+1) dimensions and the extra 6 dimensions,
which is the most general structure compatible with 
the (3+1)-dimensional 
Lorentz symmetry \cite{Nishimura:2013moa}.
Our ultimate goal is 
to determine the solution that describes our real world
by identifying a specific solution that appears
from the dominant configurations 
obtained by numerical simulations.
We consider that some, if not all, of
the qualitative features of such a solution
should be 
captured by the solutions generated by our method.

In typical solutions, we find that the (3+1)-dimensional 
space--time exhibits an expanding behavior with a smooth structure.
This is in sharp contrast to
the situation with the previous Monte Carlo studies of the Lorentzian
type IIB matrix model \cite{Aoki:2019tby},
where it was found that the 3-dimensional expanding
space is described essentially by the Pauli matrices.
This singular structure has been attributed to the approximation 
used to avoid the sign problem in the Monte Carlo simulation.
Indeed a clear departure from the Pauli-matrix structure 
has been observed in Ref.~\cite{Nishimura:2019qal},
where the sign problem was treated correctly
by the complex Langevin method for a simpler bosonic Lorentzian model.
Based on this observation, 
it has been conjectured that a smooth (3+1)-dimensional 
expanding space--time 
should emerge dynamically 
from the Lorentzian type IIB matrix model
in the large-$N$ limit. 
Our results for the classical solutions support this conjecture.

Another issue we would like to address in this paper concerns how
the Standard Model particles
emerge from the type IIB matrix model
at low energy.
%
In fact,
Refs.~\cite{Chatzistavrakidis:2011gs,Steinacker:2014fja,Aoki:2014cya}
discussed matrix configurations representing intersecting D-branes
that can accommodate
Standard Model fermions although it was not clear whether such configurations
can be realized as classical solutions.
%
Here 
we find that a slight extension of
the simple direct-product structure,
which we refer to as the ``quasi'' direct-product structure above,
naturally realizes
classical solutions
with a block-diagonal structure 
in the extra dimensions, which corresponds to intersecting D-branes.
By choosing the dimensionality of the intersecting D-branes 
appropriately,
we observe a trend that 
the Dirac operator acquires a 
zero mode
in the large-$N$ limit.
We also confirm that the wave function of the zero mode is
localized at a point, which is consistent with the picture of 
intersecting D-branes.

In this scenario, the Dirac zero modes appear in pairs with 
opposite chirality.
As simple ways to obtain chiral fermions
from classical solutions in the type IIB matrix model,
one can either compactify the model on a noncommutative 
torus \cite{Aoki:2008ik,Aoki:2010gv}
or one can replace the matrices by operators \cite{Honda:2019bdi}
assuming $N=\infty$.
However, it 
is still an open question
whether the original type IIB matrix model
can give rise to chiral fermions at low energy
starting from finite $N$ and taking the large-$N$ limit.

The rest of this paper is organized as follows.
In Sect.~\ref{sec:classicalsolution}, we 
explain how we generate
classical solutions of the type IIB matrix model
assuming a quasi-direct-product 
structure for the (3+1) dimensions and the extra 6 dimensions.
In Sect.~\ref{Sec:analysis_X}, we investigate the space--time structure 
in the (3+1) dimensions and show that the expanding behavior
with a smooth structure appears in typical solutions.
In Sect.~\ref{sec: results_chiralzeromode}, we show that
the block-diagonal structure in the extra dimensions 
that appears naturally in typical solutions
can be viewed as a realization of 
intersecting D-branes.
In particular, we show that the Dirac zero modes appear
with localized wave functions when the dimensionality of the D-branes
is chosen appropriately.
Section \ref{conlusion} is devoted to a summary and discussions.
%
\section{Generating classical solutions}
\label{sec:classicalsolution}
\setcounter{equation}{0}

In this section, after a brief review
of the type IIB matrix model,
we explain how we generate its classical solutions numerically.

\subsection{The type IIB matrix model and its classical solutions} 

The action of the type IIB matrix model is given 
as \cite{Ishibashi:1996xs}
\begin{align}
S&=S_\mathrm{b} + S_\mathrm{f}\ , 
\label{S_tot}
\\
S_\mathrm{b} &= -\frac{1}{4g^2} 
\Tr \paren{\comm{A^M}{A^N} \comm{A_M}{A_N}} \ , 
\label{S_b}
\\
S_\mathrm{f} &= -\frac{1}{2g^2} 
\Tr \paren{\Psi_{\alpha} (\mathcal{C} \Gamma^M)_{\alpha\beta} 
\comm{A_M}{\Psi_{\beta}}} \ ,
\label{S_f}
\end{align}
where $A^M \ (M=0,\ldots,9)$ and $\Psi_{\alpha}\ (\alpha=1,\ldots,16)$ 
are  $N \times N$ traceless Hermitian matrices, while
$\Gamma^M$ and $\mathcal{C}$ 
are the gamma matrices and the charge conjugation matrix 
in 10 dimensions after the Weyl projection.
The indices $M$ and $N$ are raised and lowered by using the
metric $\eta={\rm diag}(-1,1, \ldots , 1)$.
The action is invariant under the SO(9,1) transformation,
where $A^M$ and $\Psi_{\alpha}$ transform as a Lorentz vector 
and a Majorana--Weyl spinor, respectively.
The action is also invariant under the SU$(N)$ transformation
\begin{equation}
\label{eq: SUN_transformation}
A_M  \mapsto  U A_M U^\dagger \ , \quad
\Psi_\alpha  \mapsto U \Psi_\alpha U^\dagger \ ,
\end{equation}
where $U \in \mathrm{SU}(N)$.
The Euclidean version can be defined by the replacement 
$A_0 = i A_{10}$, where $A_{10}$ is Hermitian,
but here we stick to the Lorentzian version given above.
In order to make the Lorentzian model well-defined, 
we impose the constraints
\begin{equation}
\label{eq: IRcutoff}
\frac{1}{N}\Tr (A_0)^2 = 
\kappa \ , \quad \frac{1}{N}\Tr (A_I)^2 = 1 \quad (I=1,\ldots,9) \ ,
\end{equation}
corresponding to the IR cutoffs introduced in Ref.~\cite{Kim:2011cr}.\footnote{Strictly speaking, 
the IR cutoffs introduced in Ref.~\cite{Kim:2011cr} take the 
form of inequalities unlike Eq.~(\ref{eq: IRcutoff}). 
This is not a big difference, however, since the inequalities
are actually saturated in numerical simulations for entropic reasons.}

At late times, we expect that
the classical approximation is valid due to the expansion of space.
We therefore solve the classical equation of motion of this model,
which reads
\begin{equation}
\label{eq: EOM}
\comm{A^M}{\comm{A_M}{A_0}}-\xi A_0 = 0 \ ,\quad
\comm{A^M}{\comm{A_M}{A_I}}-\zeta A_I = 0 \ ,
\end{equation}
where $\xi$ and $\zeta$ are the Lagrange multipliers
corresponding to the constraints \eqref{eq: IRcutoff}.
Note here that if $\xi=\zeta=0$,
classical solutions are always simultaneously diagonalizable, 
as is proved in Appendix A of 
Ref.~\cite{Steinacker:2017vqw}.
Thus the existence of the constraints (\ref{eq: IRcutoff})
is crucial in obtaining
nontrivial solutions.

In this paper we search for solutions 
with a quasi-direct-product structure 
for the (3+1) dimensions and 
the extra 6 dimensions \cite{Nishimura:2013moa}
given as
\begin{equation}
\label{eq: ansatz of A}
A_\mu = X_\mu \otimes M \quad (\mu=0,1,2,3) \ , \quad
A_a = \mathbf{1} \otimes Y_a \quad (a=4, \ldots, 9) \ ,
\end{equation}
where the $N_{\rm X} \times N_{\rm X}$ Hermitian matrices $X_\mu$ represent the 
structure in the (3+1) dimensions,
and the $N_{\rm Y} \times N_{\rm Y}$ Hermitian matrices $Y_a$ represent the 
structure in the extra 6 dimensions with $N=N_{\rm X} N_{\rm Y}$.
This structure 
is actually compatible with the SO(3,1) symmetry
\begin{equation}
\label{eq: Lorentzsymmetry}
U A_\mu U^\dagger = {O_\mu}^\nu A_\nu\ ,\quad
U A_a U^\dagger = A_a  \ ,
\end{equation}
where $U= g \otimes \mathbf{1} \in \mathrm{SU}(N)$,
$g \in \mathrm{SU}(N_{\rm X})$ and 
${O_\mu}^\nu \in \mathrm{SO}(3,1)$
if $X_\mu$ satisfy $g X_\mu g^\dagger = {O_\mu}^\nu X_\nu$.
Note that 
the $N_{\rm Y} \times N_{\rm Y}$ Hermitian matrix $M$ does not have to be
$M = \mathbf{1}$, which would correspond 
to the direct-product structure.
We will see that the matrix $M\neq \mathbf{1}$ 
plays an important role in 
making $Y_a$ block-diagonal, which will be interpreted later as
dynamical generation of intersecting D-branes.

\subsection{The algorithm}

Since time does not exist \emph{a priori} in the type IIB matrix model, 
solving the classical equation of motion (\ref{eq: EOM})
requires some method other than just solving differential equations
unlike in ordinary classical dynamics.
Here we define a ``potential''
\begin{equation}
V = \Tr \paren{\comm{A^M}{\comm{A_M}{A_0}} - \xi A_0}^2 
+ \Tr \paren{\comm{A^M}{\comm{A_M}{A_I}} - \zeta A_I}^2 \ ,
\end{equation}
which takes the minimum $V=0$ if and only if
$A_M$ solves the classical equation of motion (\ref{eq: EOM}).
By minimizing $V$ using the gradient descent method
starting from random configurations,
we can generate classical solutions.
Within our ansatz \eqref{eq: ansatz of A},
the constraints \eqref{eq: IRcutoff} reduce to
\begin{align}
\label{eq:normalization}
\frac{1}{N_{\rm X}}\mathrm{Tr} \paren{X_0}^{2} = \kappa \ , 
\quad \frac{1}{N_{\rm Y}}\mathrm{Tr}M^{2} & =1 \ , \\
\frac{1}{N_{\rm X}}\sum_{i=1}^{3}
\mathrm{Tr}\paren{X_i}^{2}+\frac{1}{N_{\rm Y}}
\sum_{a=4}^{9}\mathrm{Tr}\paren{Y_a}^{2} & =1 \ ,
\end{align}
where Eq.~(\ref{eq:normalization})
is chosen by using
the redundancy under the rescaling
$X_\mu \mapsto s X_\mu$ and $M \mapsto s^{-1} M$,
which leaves $A_\mu = X_\mu \otimes M$ unaltered.

The initial configuration is prepared 
by generating Gaussian variables for each element
of $X_\mu$, $Y_a$ and $M$ in Eq.~(\ref{eq: ansatz of A}).
Since the equation of motion (\ref{eq: EOM})
is invariant under rescaling 
\begin{equation}
A_M \to s A_M \ , \quad \xi \to s^2 \xi \ ,
\quad \zeta \to s^2 \zeta \ ,
\label{rescaleing_XYM}
\end{equation}
we make this rescaling
at each step of the algorithm in such a way that
the second equation of \eqref{eq: IRcutoff} is satisfied.
%
The algorithm proceeds by
updating $X_\mu$, $Y_a$ and $M$
with the increment
\begin{equation}
\delta X_\mu = -\epsilon \pdv{V}{X^{*}_\mu}\ ,\quad
\delta Y_a = -\epsilon \pdv{V}{Y^{*}_a}\ ,\quad
\delta M = -\epsilon \pdv{V}{M^{*}}\ ,
\label{updating_XYM}
\end{equation}
where the stepsize $\epsilon$ is taken to be 
small enough to make sure that
$V$ decreases at each step. 
We repeat Eqs.~(\ref{rescaleing_XYM}) and (\ref{updating_XYM})
until $V$ becomes as small as $O(10^{-4})$.
The first equation of Eq.~(\ref{eq:normalization}) 
is not imposed on configurations during this procedure,
and we use it only to determine the value of $\kappa$ after
a classical solution is found.
In this way, we can obtain infinitely many solutions 
with some $\kappa$
depending on the initial random configuration and the initial 
values of $\xi$ and $\zeta$. 

\subsection{Typical solutions}
\label{sec:typical-sol}

In typical solutions generated by our algorithm,
it turns out that the matrices $M$ and $Y_a$ take 
the form
\begin{equation}
\label{eq: correspondence_MY}
M= \sqrt{\frac{N_{\rm Y}}{N_{\rm Y} - n_0 }}
\begin{pmatrix}
-\mathbf{1}_{n_{-}} &     & \\
    & \mathbf{0}_{n_0} & \\
    &      & \mathbf{1}_{n_{+}}
\end{pmatrix} \ , \quad
Y_a=
\begin{pmatrix}
Y_a^{(-)} &     & \\
    & Y_a^{(0)} & \\
    &      & Y_a^{(+)}
\end{pmatrix}
\end{equation}
up to the SU($N_{\rm Y}$) symmetry
and $N_{\rm Y}=n_{-} + n_0 + n_{+}$.\footnote{The coefficient 
$N_{\rm Y}/(N_{\rm Y} - n_0)$
for $M$ is simply due to the 
chosen normalization $(1/N_{\rm Y})\mathrm{Tr}M^{2} =1$.}
This implies that
\begin{align}
\frac{N_{\rm Y}-n_0}{N_{\rm Y}} M^3 =M \ , \quad \quad \quad
\label{eq: M} 
\comm{M}{Y_a} =0 \ .
\end{align}
Plugging Eq.~(\ref{eq: M})
into the equation of motion (\ref{eq: EOM})
with the ansatz \eqref{eq: ansatz of A},
we find that $X_\mu$ and $Y_a$ decouple as
\begin{align}
\comm{X^\nu}{\comm{X_\nu}{X_0}} - \xi X_0 &= 0  \ , \nonumber \\
\comm{X^\nu}{\comm{X_\nu}{X_i}} - \zeta X_i &= 0 \quad (i=1,2,3) \ ,
\label{eq: X_mu}
\\
\label{eq: Y_a}
\comm{Y^b}{\comm{Y_b}{Y_a}} - \zeta Y_a&= 0  \quad (a=4, \ldots , 9) \ . 
\end{align}

In this work,
we set the parameters $\xi$ and $\zeta$ to be equal 
in order to respect the SO(3,1) symmetry
and take them to be positive.
Note that 
once we impose this condition on the initial values of $\xi$ and $\zeta$,
it is satisfied automatically at each step.
The value of $\kappa$ turns out to be close to $1/3$ as far as 
we use the same Gaussian distribution for all the elements of 
$X_\mu$, $Y_a$ and $M$ in the initial configuration.
A solution that belongs to the above class 
(\ref{eq: correspondence_MY})
is obtained without being 
trapped in a local minimum
when the value of the parameter $\zeta$ lies
within the range
\begin{equation}
\label{eq:mass_region} 
0.1 \lesssim \zeta \lesssim 0.5 \ .
\end{equation}
%


\section{Space-time structure in the (3+1) dimensions}
\label{Sec:analysis_X}
\setcounter{equation}{0}

%

Using the method described in the previous section, 
we obtain classical solutions of the type IIB matrix model
within the ansatz (\ref{eq: ansatz of A}).
By generating the initial configurations randomly,
we can obtain as many solutions as we wish
satisfying \eqref{eq: M}--\eqref{eq: Y_a}.
In this section, we focus on 
the matrices $X_{\mu}$ in Eq.~(\ref{eq: ansatz of A}),
which represent the structure of the (3+1)-dimensional space--time.

\subsection{Band-diagonal structure}
\label{SS:Config_X}

\begin{figure}
\centering
\includegraphics[width=10cm]{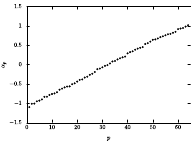}
\caption{The eigenvalues $\alpha_{p} \ (p=1,\ldots,N_{\rm X})$ of 
$X_{0}$ are plotted against the label $p$
in the ascending order
for a typical solution with $N_{\rm X}=64$. 
}
\label{fig:X0}
\end{figure}

Using the SU($N$) symmetry \eqref{eq: SUN_transformation} 
with $U=g \otimes \mathbf{1}$, $g \in \mathrm{SU}(N_{\rm X})$, 
we can choose a basis in which $X_{0}$ is diagonal 
with its elements arranged in ascending order.
In Fig.~\ref{fig:X0}, we show the eigenvalues 
$\alpha_{p}\ (p=1,2,\ldots,N_{\rm X})$ of $X_0$ for a
typical solution with $N_{\rm X} = 64$. 
We find that the eigenvalues are almost uniformly distributed,
which suggests that we can define a continuous time 
in the $N_{\rm X} \to \infty$ limit.

\begin{figure}
\centering
\includegraphics[width=10cm]{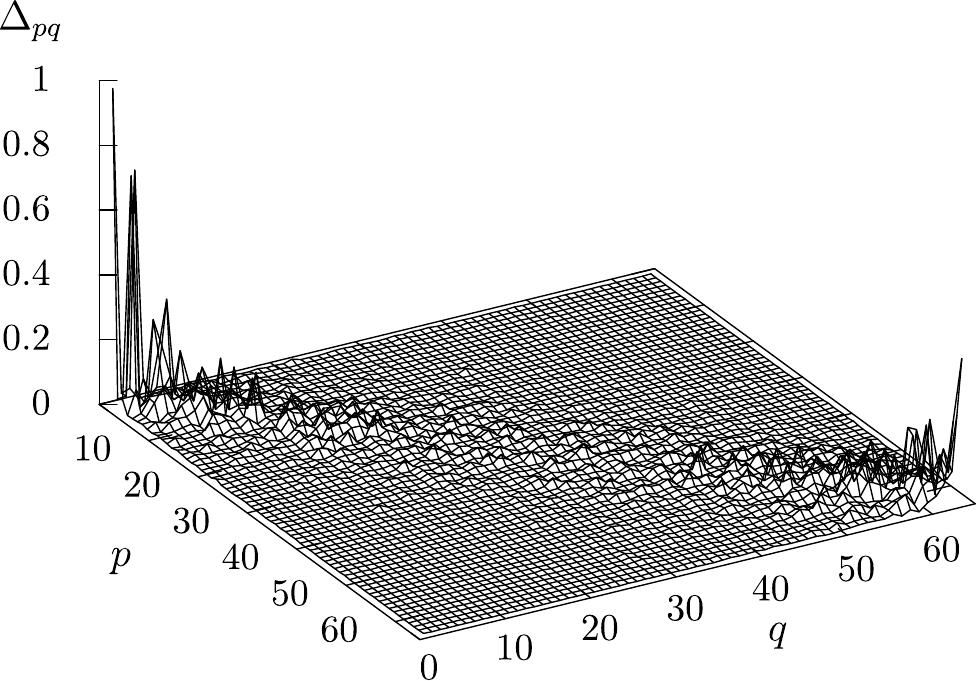}
\caption{The quantity $\Delta_{pq}$
defined by Eq.~(\ref{eq:ampX})
is plotted against the labels $p$ and $q$
for the typical solution with $N_{\rm X}=64$. 
}
\label{fig:ampX}
\end{figure}

In the basis that makes $X_{0}$ diagonal,
the spatial matrices $X_{i} \  (i=1,2,3)$ are not diagonal in general,
but they actually turn out to be band-diagonal.
In order to see it, we plot
\begin{equation}
\label{eq:ampX}
\Delta_{pq} 
=\sum_{i=1}^{3}\left|(X_{i})_{pq}\right|^{2} 
\end{equation}
for the same solution in Fig.~\ref{fig:ampX}.
We find that
$\Delta_{pq}$ becomes
very small for $|p-q|\geq n$ with some integer $n$,
which is $n \sim 10$ for the configuration in Fig.~\ref{fig:ampX}.
The band-diagonal structure guarantees the locality of time, 
which enables us to extract the time evolution as we discuss
in the next section. 

The uniform distribution of $\alpha_p$ and 
the band-diagonal structure in $X_i$ are common features
of the solutions we obtain.
They are actually shared also by the dominant configurations 
generated by previous Monte Carlo 
studies\cite{Ito:2013ywa,Ito:2015mxa,Ito:2017rcr,Azuma:2017dcb}.
As we mentioned 
below (\ref{eq: EOM}),
the classical solutions are strictly diagonal for $\xi=\zeta=0$.
Therefore, the band-diagonal structure may be viewed
as a deviation from the diagonal matrix due to 
finite $\xi$ and $\zeta$.

\subsection{Extracting the time evolution}
\label{SS:Time_evolution}

Next we discuss how we extract the time evolution
from a given classical solution. 
The band-diagonal structure discussed in the previous section
plays a crucial role here.

Let us define an $n\times n$ block $\bar{X}_i(t)$ by
\begin{equation}
(\bar{X}_i (t))_{rs}
= (X_{i})_{k+r, k+s}
\quad (r, s=1,2,\ldots,n) \ ,
\end{equation}
where $k=0,1,\ldots,N_{\rm X}-n$
and the argument $t$ represents the time defined by
\begin{equation}
t
=\frac{1}{n}\sum_{r=1}^{n}\alpha_{k+r}  \ .
\end{equation}
This block matrix $\bar{X}_i(t)$ represents the space structure
at time $t$, from which one can obtain
the time evolution of the 3-dimensional space.
%

\begin{figure}
\centering
\includegraphics[width=10cm]{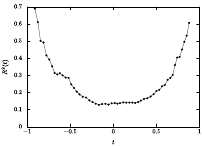}
\caption{The extent of space $R^{2}(t)$ is plotted 
against $t$ for the typical solution with $N_{\rm X}=64$
and the block size $n=10$. 
}
\label{fig:R2}
\end{figure}

For instance, we define
the extent of space at $t$ by
\begin{equation}
R^{2}\left(t\right)
=\frac{1}{n}\sum_{i=1}^{3}
\mathrm{tr}\left(\bar{X}_{i} (t)\right)^2 \ ,
\end{equation}
where the symbol
``$\tr$'' represents the trace over an $n \times n$ matrix.
In Fig.~\ref{fig:R2}, we plot $R^{2}(t)$ for 
the typical solution discussed in Sect.~\ref{SS:Config_X}. 
We find that $R^{2}(t)$ is roughly symmetric under $t\mapsto -t$
reflecting the symmetry of the model under $A_0 \mapsto - A_0$. 
In particular, an expanding behavior is seen at $t>0$,
which is consistent with the results of 
simulations \cite{Nishimura:2019qal} at late time.


\begin{figure}
\centering
\includegraphics[width=10cm]{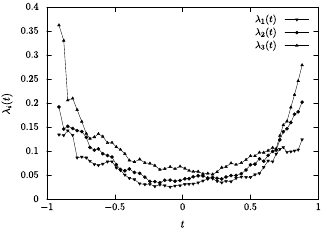}
\caption{The eigenvalues $\lambda_i (t)$ 
of $T_{ij}(t) \ (i,j=1,2,3)$ 
are plotted against $t$ 
for the typical solution with $N_{\rm X}=64$
and the block size $n=10$. 
}
\label{fig:T}
\end{figure}

Next we define a $3\times3$ 
real symmetric tensor
\begin{equation}
T_{ij}\left(t\right)=\frac{1}{n}\mathrm{tr}
\left(\bar{X}_{i}\left(t\right)\bar{X}_{j}\left(t\right)\right) \ ,
\end{equation}
whose
eigenvalues $\lambda_i(t)$,
which we order as 
$\lambda_{1}(t) \leq \lambda_{2}(t) \leq \lambda_{3}(t)$,
represent how the 3-dimensional space extends in each direction.
Note that they are related to the extent of space $R^2(t)$ as
\begin{equation}
R^{2}\left(t\right)=\sum_{i=1}^{3} \lambda_{i} (t)  \ .
\end{equation}
In Fig.~\ref{fig:T}, we plot
the eigenvalues $\lambda_i(t)$
for the typical solution discussed above.
The three eigenvalues $\lambda_i(t)$
turn out to be
roughly equal at each time.
We have also confirmed that the three eigenvalues
tend to come closer to each other
as one increases the matrix size $N_{\rm X}$,
suggesting that the SO(3) symmetry is recovered
in the limit of infinite matrix size.


\begin{figure}
\centering
\includegraphics[width=10cm]{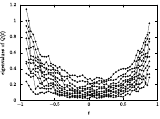}
\caption{The eigenvalues of $Q(t)$ are plotted against 
$t$ for the typical solution with $N_{\rm X}=64$.
There are ten of them reflecting the chosen block size $n=10$. 
}
\label{fig:Q}
\end{figure}


In order to probe
the structure of the 3-dimensional space 
emerging from
the classical solutions in more detail,
we define an $n \times n$ matrix
\begin{equation}
Q\left(t\right)=\sum_{i=1}^{3} \bar{X}_{i}(t)^2  \ ,
\end{equation}
whose 
eigenvalues $q_{r}(t)$
represent 
the radial distribution of the points which describe the 3-dimensional space. 
Note that the eigenvalues $q_{r}(t)$ 
are related to the extent of space $R^{2}(t)$ as
\begin{equation}
R^{2} (t) 
=\frac{1}{n}\sum_{r=1}^{n} q_{r} (t) \ .
\end{equation}
In Fig.~\ref{fig:Q}, we plot the eigenvalues of $Q(t)$ for the same typical
solution discussed above.
We find that they are densely distributed
at each time
and that all the eigenvalues are growing with time
in the $t>0$ region.
This is in sharp contrast to
the results for the configurations obtained by
the previous Monte Carlo studies of the Lorentzian
type IIB matrix model \cite{Aoki:2019tby},
where only two of the eigenvalues 
grow with time,
while the others remain small and constant.
As we mentioned in the Introduction,
the latter behavior is interpreted as a consequence
of the approximation used
in the Monte Carlo studies to avoid the sign problem.
Our results for the classical solutions 
suggest the emergence of \emph{smooth}
(3+1)-dimensional expanding space--time in the large-$N$ limit.

\section{The emergence of intersecting D-branes}
\label{sec: results_chiralzeromode}
\setcounter{equation}{0}

In this section we discuss how chiral zero modes
appear from the type IIB matrix model.
A general discussion on the Dirac equation
implies that it is necessary to have zero modes in
the 6-dimensional Dirac operator.
In fact, we show that
the block-diagonal structure (\ref{eq: correspondence_MY})
of the typical solutions
can be regarded as the emergence of intersecting D-branes.
Indeed we find that zero modes appear in the limit 
of infinite matrix size if the dimensionality of the branes is 
chosen appropriately.

\subsection{The Dirac equation}


From the fermionic action (\ref{S_f}) of the type IIB matrix model, 
one obtains the 10-dimensional Dirac equation
\begin{equation}
\label{eq: 10d_Dirac}
\Gamma^M \comm{A_M}{\Psi}=0 \ ,
\end{equation}
where $\Psi$ satisfies the Weyl condition 
\begin{equation}
\label{eq: chirality_in_10d}
\Gamma_\chi \Psi = \Psi
\end{equation}
with the chirality operator $\Gamma_\chi$ in 10 dimensions.
Here we decompose the 10-dimensional gamma matrices $\Gamma^M$ 
as
\begin{equation} 
\Gamma^\mu= \rho^\mu \otimes \mathbf{1} \ , \quad
\Gamma^a= \rho_\chi \otimes \gamma^a \ ,
\end{equation} 
where $\rho^\mu$ and $\gamma^a$ 
are the 4-dimensional and 6-dimensional gamma matrices, respectively.
Note that the chirality operator $\Gamma_\chi$ is also decomposed as
\begin{equation}
\Gamma_\chi = \rho_\chi \otimes \gamma_\chi \ ,
\end{equation}
where $\rho_\chi$ and $\gamma_\chi$ are the chirality operators 
in 4 and 6 dimensions, respectively.

Suppose $\Psi$ is chiral in 4 dimensions, meaning
\begin{equation}
\label{eq: chirality_in_4d_Psi}
(\rho_\chi \otimes \mathbf{1}) \Psi = \pm \Psi \ .
\end{equation}
Due to the chirality (\ref{eq: chirality_in_10d}) in 10 dimensions,
Eq.~(\ref{eq: chirality_in_4d_Psi}) implies
\begin{equation}
\label{eq: chirality_in_6d_Psi}
\paren{\mathbf{1} \otimes \gamma_\chi} \Psi = \pm \Psi \ ;
\end{equation}
i.e., $\Psi$ is chiral also in 6 dimensions.

Let us next decompose Eq.~(\ref{eq: 10d_Dirac}) into two terms as
\begin{equation}
\Gamma^\mu \comm{A_\mu}{\Psi}+ \Gamma^a \comm{A_a}{\Psi}=0 \ .
\label{eq:ac-decompose}
\end{equation}
Note that the first term and the second term have opposite
chirality in 4 dimensions as well as in 6 dimensions,
which implies that each term has to vanish separately.
Thus, in order to obtain 
chiral fermions in four dimensions,
we need to have Dirac zero modes in 6 dimensions; i.e.,
\begin{equation}
\Gamma^a \comm{A_a}{\Psi}=0 \ .
\label{Dirac-eq-6d}
\end{equation}

Let us now consider that $A_M$ is a classical solution with the 
quasi-direct-product structure (\ref{eq: ansatz of A}).
Since $A_a = \mathbf{1} \otimes Y_a$,
the general solution to Eq.~(\ref{Dirac-eq-6d}) can be obtained
by decomposing $\Psi$ as
\begin{equation}
\Psi = \psi^\mathrm{(4d)} \otimes \psi^\mathrm{(6d)} \ , 
\end{equation}
where the 4-dimensional and 6-dimensional 
gamma matrices act only on $\psi^\mathrm{(4d)}$
and $\psi^\mathrm{(6d)}$, respectively.
Thus, in order to satisfy Eq.~(\ref{Dirac-eq-6d}), we only need to require
\begin{equation}
\label{eq: chirality_in_6d_psi}
\gamma^a \comm{Y_a}{\psi_a^\mathrm{(6d)}}=0 \ .
\end{equation}

\subsection{Dirac zero modes from intersecting D-branes}

When the configuration of $Y_a$ takes the block-diagonal form
(\ref{eq: correspondence_MY}),
the eigenvalue problem associated with the
Dirac operator on the left-hand side of Eq.~(\ref{eq: chirality_in_6d_psi})
reduces to that for each block matrix of $\psi^\mathrm{(6d)}$.
Note that
the diagonal blocks of $Y_a$ can be interpreted as D-branes,
whereas an off-diagonal block of $\psi^\mathrm{(6d)}$
can be interpreted as the open string ending on the corresponding
two D-branes.
When the two D-branes intersect with each other at a point,
the length of the open string can become zero
as it comes close to the intersecting point,
and therefore a chiral zero mode appears there.
In what follows, we confirm that this indeed happens for
the typical solutions.

For that purpose, it suffices to consider
the solutions of \eqref{eq: Y_a}
with a block-diagonal structure
\begin{align}
\label{eq: form_Ya}
Y_a &=
\begin{pmatrix}
Y_a^{(1)} & \\
 & Y_a^{(2)}
\end{pmatrix} \ ,
\end{align} 
where the block matrices
$Y_a^{(1)}$ and $Y_a^{(2)}$ have to satisfy
\eqref{eq: Y_a} with the same $\zeta$,
which we set to $\zeta=1$ without loss of generality.
We denote the size of the block matrices $Y_a^{(1)}$ and $Y_a^{(2)}$
as $N_{\rm Y}^{(1)}$ and $N_{\rm Y}^{(2)}$, respectively.

Corresponding to the block-diagonal structure (\ref{eq: form_Ya})
of $Y_a$,
we focus on the off-diagonal block $\varphi$ in
\begin{align}
\psi^\mathrm{(6d)}&=
\begin{pmatrix}
 & \varphi \\
 & 
\end{pmatrix} \ ,
\label{eq: off-diagonal_block_psi}
\end{align} 
for which the eigenvalue problem reduces to 
\begin{equation}
\label{eq: Diraceq_6d}
\gamma^a_{\alpha \beta} \boxbc{Y_a^{(1)} \varphi_\beta
- \varphi_\beta Y_a^{(2)}} =\lambda \varphi_\alpha \ ,
\end{equation}
where $\varphi_\alpha \ (\alpha=1,\ldots,8)$ are the eigenvectors.
The eigenvalue $\lambda$ corresponds to the mass in the (3+1) dimensions
as one can see from Eq.~(\ref{eq:ac-decompose}).

We obtain numerically the eigenvalue spectrum of the Dirac operator
for various pairs of classical solutions
$Y_a^{(1)}$ and $Y_a^{(2)}$
to see in which case we obtain zero modes in the limit of
infinite matrix size.
Also, we look into the structure of the 
eigenvectors $\varphi_\alpha$, which we call the ``wave functions''
in what follows, and see whether it is localized
as is expected from the picture of intersecting D-branes.

Since the Dirac operator under consideration anti-commutes
with $\gamma_\chi$,
the eigenvalues $\lambda$ and $- \lambda$ appear in pairs.
If $\varphi$ is an eigenvector corresponding to the
eigenvalue $\lambda$,
then $\gamma_{\chi}\varphi$ is an eigenvector corresponding to the
eigenvalue $- \lambda$.
For the zero modes corresponding to $\lambda=0$,
one can choose them to have definite chirality
$\gamma_\chi \varphi = \pm \varphi$.
In order to see the localization of the zero modes, we 
need to consider the right-handed and left-handed components separately,
which can be extracted as
\begin{equation}
\varphi_{\mathrm{R} \alpha} = 
\frac{1+ \gamma_\chi}{2} \varphi_\alpha \ , \quad
\varphi_{\mathrm{L} \alpha} = 
\frac{1- \gamma_\chi}{2} \varphi_\alpha \ .
\label{eq: unitary_transf_psi}
\end{equation}

Furthermore, we should take into account that the 
classical solution of the form Eq.~(\ref{eq: form_Ya}) has
symmetry under
\begin{equation}
\label{eq: UEX}
Y_a  \mapsto  g \, Y_a \, g^\dagger \ , 
\end{equation}
where the $g \in \mathrm{SU}(N_{\rm Y})$ is block-diagonal
\begin{equation}
g =
\begin{pmatrix}
g^{(1)} & 0 \\
0 & g^{(2)}
\end{pmatrix}
\end{equation}
with $g^{(1)} \in \mathrm{SU}(N_{\rm Y}^{(1)})$ and 
$g^{(2)} \in \mathrm{SU}(N_{\rm Y}^{(2)})$, 
which preserves the structure \eqref{eq: form_Ya}.
Under this transformation, the wave functions of the 
corresponding Dirac operator transform as
\begin{equation}
\label{eq: varphi_LR_prime}
\varphi_\alpha \mapsto 
g^{(1)} \varphi_\alpha g^{(2)\dagger}  \ .
\end{equation}
We choose $g^{(1)}$ and $g^{(2)}$ 
in \eqref{eq: varphi_LR_prime} in such a way that
$\varphi_{\mathrm{R}1}$ is diagonal with positive elements
in the descending order.
This is nothing but the singular value decomposition
of $\varphi_{\mathrm{R}1}$, where
the diagonal elements correspond to the singular values.

In what follows, we change the index $a$ of $Y_a$
from $a=4,\ldots,9$ to $a=1,\ldots,6$.
From the picture of the intersecting D-branes,
we expect the emergence of Dirac zero modes
when the two branes intersect at a point.
This requires generically that the dimensionality of the branes
adds up to six.
If the sum of the dimensionality is less than six,
the two branes do not intersect, and
if the sum is more than six, 
the two branes intersect but not at a point.
In order to specify the dimensionality of the brane,
say to $d$, 
we set $6-d$ components of $Y_a$ to zero.
Under these conditions, we generate 3d and 4d solutions
of \eqref{eq: Y_a} with $\zeta=1$
numerically, and use them as the block matrices
in \eqref{eq: form_Ya} in the following analysis.
The adopted algorithm is the same as the one described in
Sect.~\ref{sec:classicalsolution} except that only
the $Y_a$ matrices are involved in our calculation here
in order to reduce the computational cost.

The 2d solutions, on the other hand,
can be constructed analytically
by defining $Z$ as
\begin{equation}
\label{eq: Z}
\comm{Y_1}{Y_2}=i Z \ ,
\end{equation}
so that we can reduce \eqref{eq: Y_a} with $\zeta=1$ to
\begin{equation}
\label{eq: Y1Z_Y2Z}
\comm{Y_2}{Z} = i  Y_1  \ , \quad
\comm{Z}{Y_1}= i  Y_2 \ .
\end{equation}
Eqs.~\eqref{eq: Z} and \eqref{eq: Y1Z_Y2Z} 
imply
\begin{equation}
Y_1 = L_1 \ , \quad Y_2 = L_2 \ , \quad Z = L_3 \ ,
\label{2d-solution-generator}
\end{equation} 
where $L_i\ (i=1,2,3)$ are some representation of the SU(2) algebra.
Here we restrict ourselves to the irreducible representation 
without loss of generality,
since the Dirac spectrum for the reducible representation 
simply becomes the sum of those for the irreducible representations
which the reducible representation decomposes to.
The above construction suggests that the 2d brane is something
like a ``fuzzy disk'', which can be obtained by projecting a
fuzzy sphere on to a plane. More precisely, it should be regarded as two
coinciding fuzzy disks corresponding to the two hemispheres
of the fuzzy sphere.


\subsection{The cases with intersection at a point}

\paragraph{``3d--3d'' ansatz}
\label{subsec: 3d-3d}
For the first brane $Y_a^{(1)}$,
we set $Y_4^{(1)} = Y_5^{(1)} = Y_6^{(1)} = 0$, 
and for the second brane $Y_a^{(2)}$,
we set $Y_1^{(2)} = Y_2^{(2)} = Y_3^{(2)} = 0$.
We use four 3d solutions for each of them.
Thus we consider $4\times 4=16$ backgrounds
with matrix sizes
$N_{\rm Y}^{(1)}=N_{\rm Y}^{(2)}=32$, $40$, $48$, $56$ and $64$,
on which we solve the eigenvalue problem (\ref{eq: Diraceq_6d}).

\begin{figure}
\centering
\includegraphics[width=10.0cm]{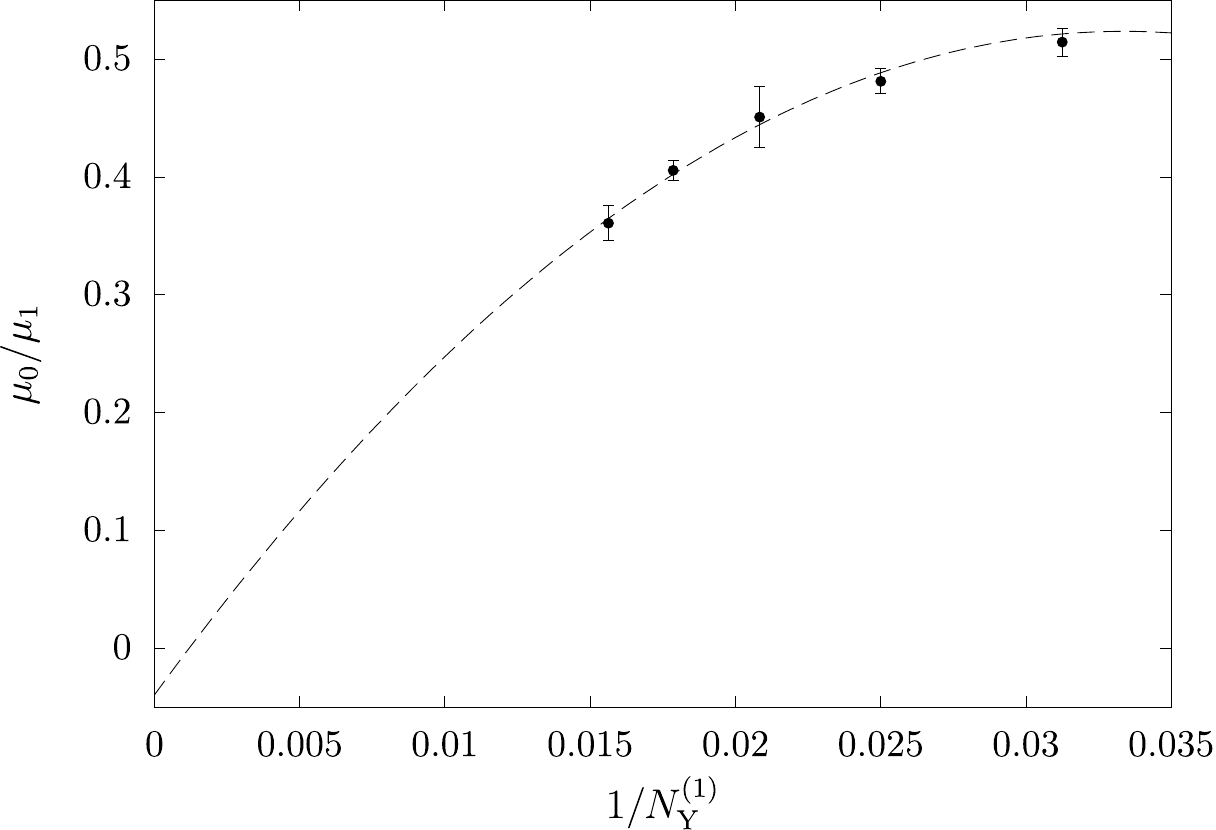}
\caption{The ratio $\mu_0/\mu_1$ 
with the ``3d--3d'' ansatz is plotted against $1/N_{\rm Y}^{(1)}$
for $N_{\rm Y}^{(1)}=N_{\rm Y}^{(2)}=32$, $40$, $48$, $56$ and $64$.
The dashed line represents a fit 
to the quadratic function 
$a+b/N_{\rm Y}^{(1)}+c/(N_{\rm Y}^{(1)})^2$ 
of $1/N_{\rm Y}^{(1)}$
with 
$a=-0.04(7)$, $b=34(6)$ and $c=-5(1) \times 10^2$.
}
\label{ratio_EV_3d}
\end{figure}

\begin{figure}
\centering
\includegraphics[width=16.5cm]{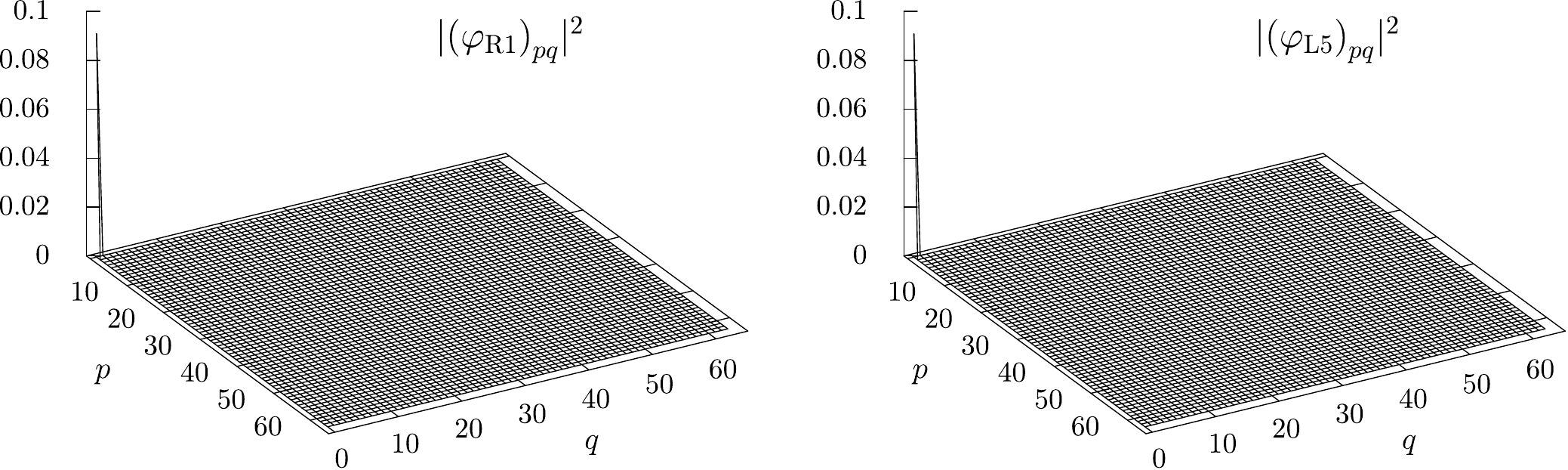}
\caption{The wave functions
$|(\varphi_{\mathrm{R}1})_{pq}|^2$ 
and $|(\varphi_{\mathrm{L}5})_{pq}|^2$
$(p=1,\ldots,N_{\rm Y}^{(1)},  \ q=1,\ldots,N_{\rm Y}^{(2)})$
with the ``3d--3d'' ansatz are plotted
for $N_{\rm Y}^{(1)}=N_{\rm Y}^{(2)}=64$.
}
\label{wavefn_3d-3d}
\end{figure}

We obtain the smallest absolute value $|\lambda|$
of the eigenvalues
for each of the 16 cases and denote the average as $\mu_0$.
Similarly, we denote as $\mu_1$ the average of  
the second smallest absolute value $|\lambda|$.
Since the normalization of the solutions 
is somewhat arbitrary for different matrix sizes,
we take the ratio $\mu_0/\mu_1$ and 
plot it against $1/N_{\rm Y}^{(1)}$
in Fig.~\ref{ratio_EV_3d}.
The fit to the quadratic function of $1/N_{\rm Y}^{(1)}$
suggests that zero modes appear in the $N_{\rm Y}^{(1)} \to \infty$ limit.

Next we consider the wave function corresponding 
to the lowest eigenvalue for one of the 16 cases with $N_{\rm Y}^{(1)}=N_{\rm Y}^{(2)}=64$.
We calculate $|(\varphi_{\mathrm{R}\alpha})_{pq}|^2$ 
for each $\alpha$
and find that the wave function almost vanishes
except for the spinor component $\alpha=1$.
Similarly, we calculate $|(\varphi_{\mathrm{L} \alpha})_{pq}|^2$ 
for each $\alpha$,
and find that the wave function almost vanishes
except for the spinor component $\alpha=5$.
In Fig.~\ref{wavefn_3d-3d},
we plot $|(\varphi_{\mathrm{R}1})_{pq}|^2$
and  $[(\varphi_{\mathrm{L}5})_{pq}|^2$, the component of the 
wave functions that has nonzero elements. 
We find that the wave functions are localized at the $(1,1)$ element.
These results are consistent with the picture that the left-handed 
and right-handed zero modes appear from the intersecting D-branes.

\paragraph{``2d--4d'' ansatz}
\label{subsec: 2d-4d}
For the first brane $Y_a^{(1)}$,
we set $Y_3^{(1)} =Y_4^{(1)} = Y_5^{(1)} = Y_6^{(1)} = 0$
and use the analytic solution corresponding to the 
irreducible representation of the SU(2) algebra
with $N_{\rm Y}^{(1)}=24$, $28$, $32$ and $36$.
For the second brane $Y_a^{(2)}$,
we set $Y_1^{(2)} = Y_2^{(2)} = 0$
and use 16 4d solutions obtained
with $N_{\rm Y}^{(2)}=(N_{\rm Y}^{(1)})^2/16$.
Thus we consider $16$ backgrounds,
on which we solve the eigenvalue problem (\ref{eq: Diraceq_6d}).\footnote{The chosen matrix size 
$N_{\rm Y}^{(2)}$ for the 4d brane
is motivated by the fact that the degrees of freedom on a lattice
with a linear extent $L$
grow as $L^2$ and $L^4$ for 2d and 4d cases, respectively.
The factor of $1/16$ in $N_{\rm Y}^{(2)}=(N_{\rm Y}^{(1)})^2/16$ is introduced
to avoid having too large $N_{\rm Y}^{(2)}$ to perform explicit calculations.}

\begin{figure}
\centering
\includegraphics[width=10.0cm]{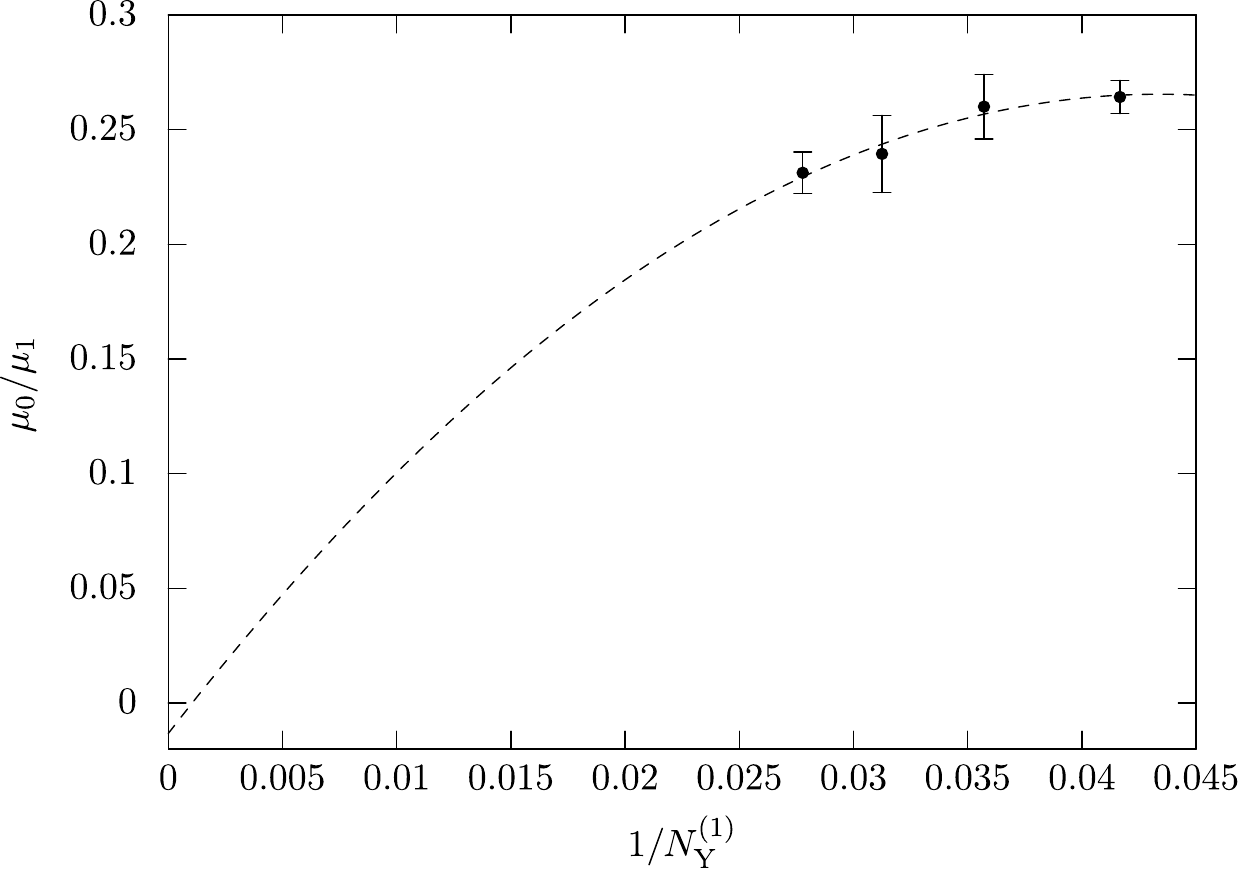}
\caption{The ratio $\mu_0/\mu_1$ 
with the ``2d--4d'' ansatz is plotted against $1/N_{\rm Y}^{(1)}$
for $N_{\rm Y}^{(1)}=24$, $28$, $32$ and $36$ with $N_{\rm Y}^{(2)}=(N_{\rm Y}^{(1)})^2/16$.
The dashed line is a fit to 
$a+ b/N_{\rm Y}^{(1)}+ c/(N_{\rm Y}^{(1)})^2$ 
with $a=-0.01(16)$, $b=13(9)$ and $c=-1.3(1.3)\times 10^2$.}
\label{ratio_EV_2d-4d_NY1=[24-28-32-36]}
\end{figure}

In the present case, the eigenvalues turn out to have two-fold
degeneracy, which may be understood from the
fact that the 2d brane is actually something like
two coinciding fuzzy disks.
(See Eq.~(\ref{2d-solution-generator}) and the lines below.)
In Fig.~\ref{ratio_EV_2d-4d_NY1=[24-28-32-36]},
we plot the ratio $\mu_0/\mu_1$
against $1/N_{\rm Y}^{(1)}$.
The fit to the quadratic function of $1/N_{\rm Y}^{(1)}$
suggests that zero modes appear in the $N_{\rm Y}^{(1)} \to \infty$ limit.

\begin{figure}
\centering
\includegraphics[width=16.5cm]{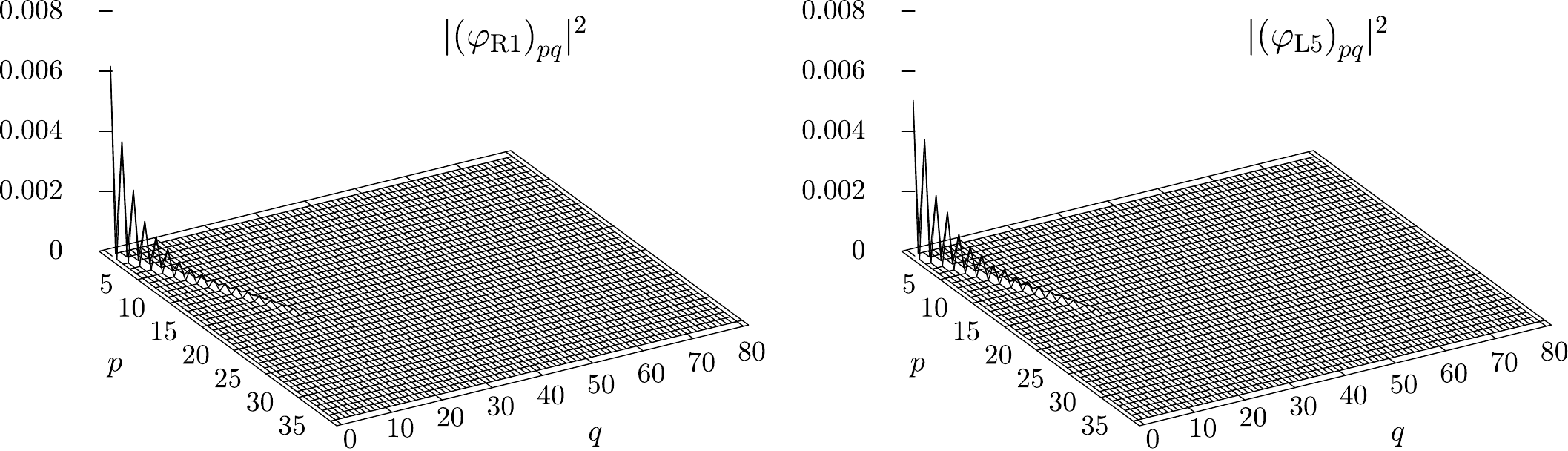}
\caption{The wave functions
$|(\varphi_{\mathrm{R}1})_{pq}|^2$ and 
$|(\varphi_{\mathrm{L}5})_{pq}|^2$
$(p=1,\ldots,N_{\rm Y}^{(1)},\ q=1,\ldots,N_{\rm Y}^{(2)})$ 
with the ``2d--4d'' ansatz are plotted
for $N_{\rm Y}^{(1)}=36$ and $N_{\rm Y}^{(2)}=81$. 
}
\label{wavefn_2d-4d}
\end{figure}

Next we consider the wave function corresponding 
to one of the 2 lowest eigenvalues
for one of the 16 cases with $N_{\rm Y}^{(1)}=N_{\rm Y}^{(2)}=64$.\footnote{The situation with
the other lowest eigenvalue is qualitatively the same. The same
comment applies also to the case with the ``2d--3d'' ansatz below.} 
We calculate $|(\varphi_{\mathrm{R}\alpha})_{pq}|^2$ 
for each $\alpha$
and find that the wave function almost vanishes
except for the spinor component $\alpha=1$.
Similarly, we calculate $|(\varphi_{\mathrm{L}\alpha})_{pq}|^2$ 
for each $\alpha$,
and find that the wave function almost vanishes
except for the spinor component $\alpha=5$.
In Fig.~\ref{wavefn_2d-4d},
we plot $|(\varphi_{\mathrm{R}1})_{pq}|^2$
and  $|(\varphi_{\mathrm{L}5})_{pq}|^2$, the component of the 
wave functions that has nonzero elements. 
We find that the wave functions are localized at the $(1,1)$ element
although
the localization is not as sharp as in the ``3d-3d'' case.
These results are consistent with the picture that the left-handed 
and right-handed zero modes appear from the intersecting D-branes.

\subsection{The cases without intersection}

\paragraph{``2d--3d'' ansatz}
\label{subsec: 2d-3d}
For the first brane $Y_a^{(1)}$,
we set $Y_3^{(1)} =Y_4^{(1)} = Y_5^{(1)} = Y_6^{(1)} = 0$
and use the analytic solution corresponding to the 
irreducible representation of the SU(2) algebra
with $N_{\rm Y}^{(1)}=32$, $48$ and $64$.
For the second brane $Y_a^{(2)}$,
we set $Y_1^{(2)} = Y_2^{(2)} =Y_3^{(2)} = 0$
and use 16 3d solutions obtained with $N_{\rm Y}^{(2)}=N_{\rm Y}^{(1)}$.
Thus we consider 16 backgrounds,
on which we solve the eigenvalue problem (\ref{eq: Diraceq_6d}).

\begin{figure}
\centering
\includegraphics[width=10.0cm]{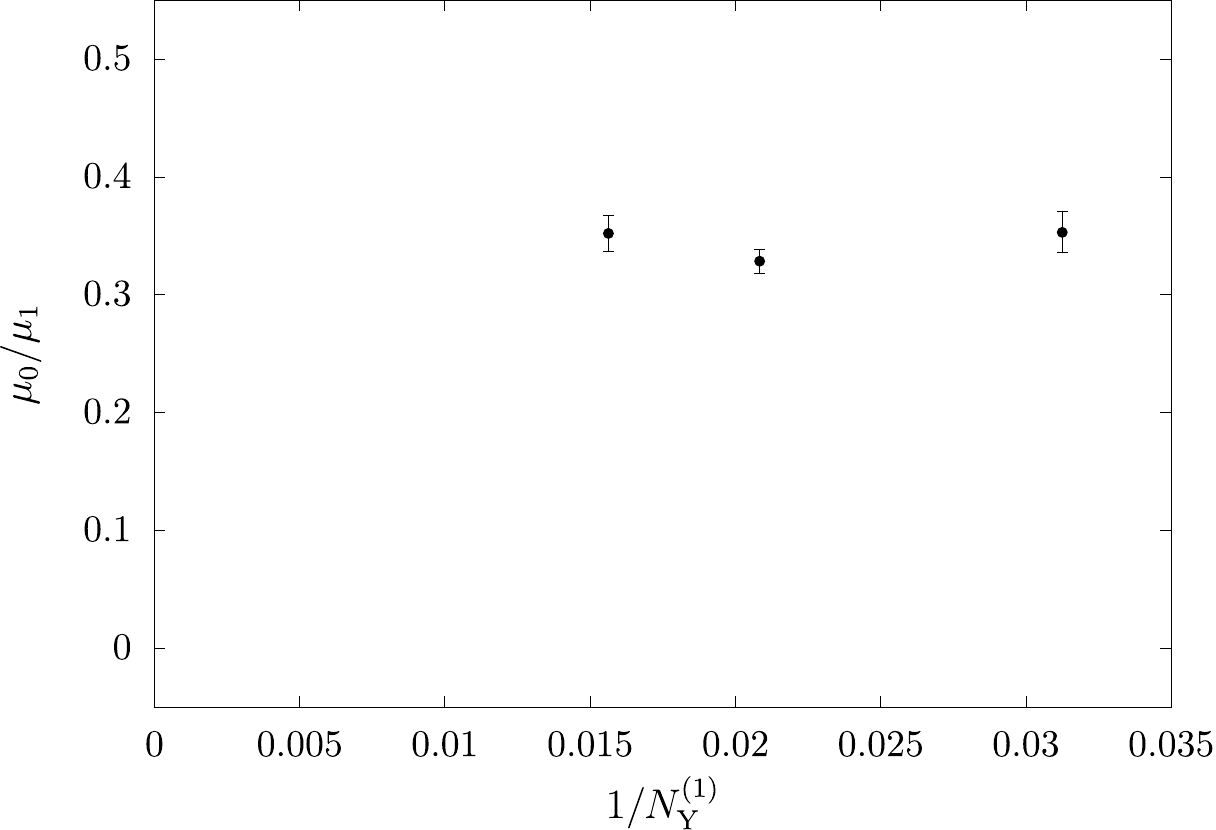}
\caption{The ratio $\mu_0/\mu_1$ 
with the ``2d--3d'' ansatz is 
plotted against $1/N_{\rm Y}^{(1)}$
for $N_{\rm Y}^{(1)}=N_{\rm Y}^{(2)}=32$, $48$ and $64$.}
\label{ratio_EV_2d-3d_NY1=[32-48-64]}
\end{figure}

As in the ``2d--4d'' case discussed above,
we have two-fold degeneracy in the eigenvalue spectrum.
In Fig.~\ref{ratio_EV_2d-3d_NY1=[32-48-64]},
we plot the ratio $\mu_0/\mu_1$
against $1/N_{\rm Y}^{(1)}$ for $N_{\rm Y}^{(1)}=N_{\rm Y}^{(2)}=32$, $48$ and $64$.
We find that it does not converge to $0$ in the 
$N_{\rm Y}^{(1)} \to \infty$ limit, which implies that
we do not obtain zero modes in this case.

\begin{figure}
\centering
\includegraphics[width=16.5cm]{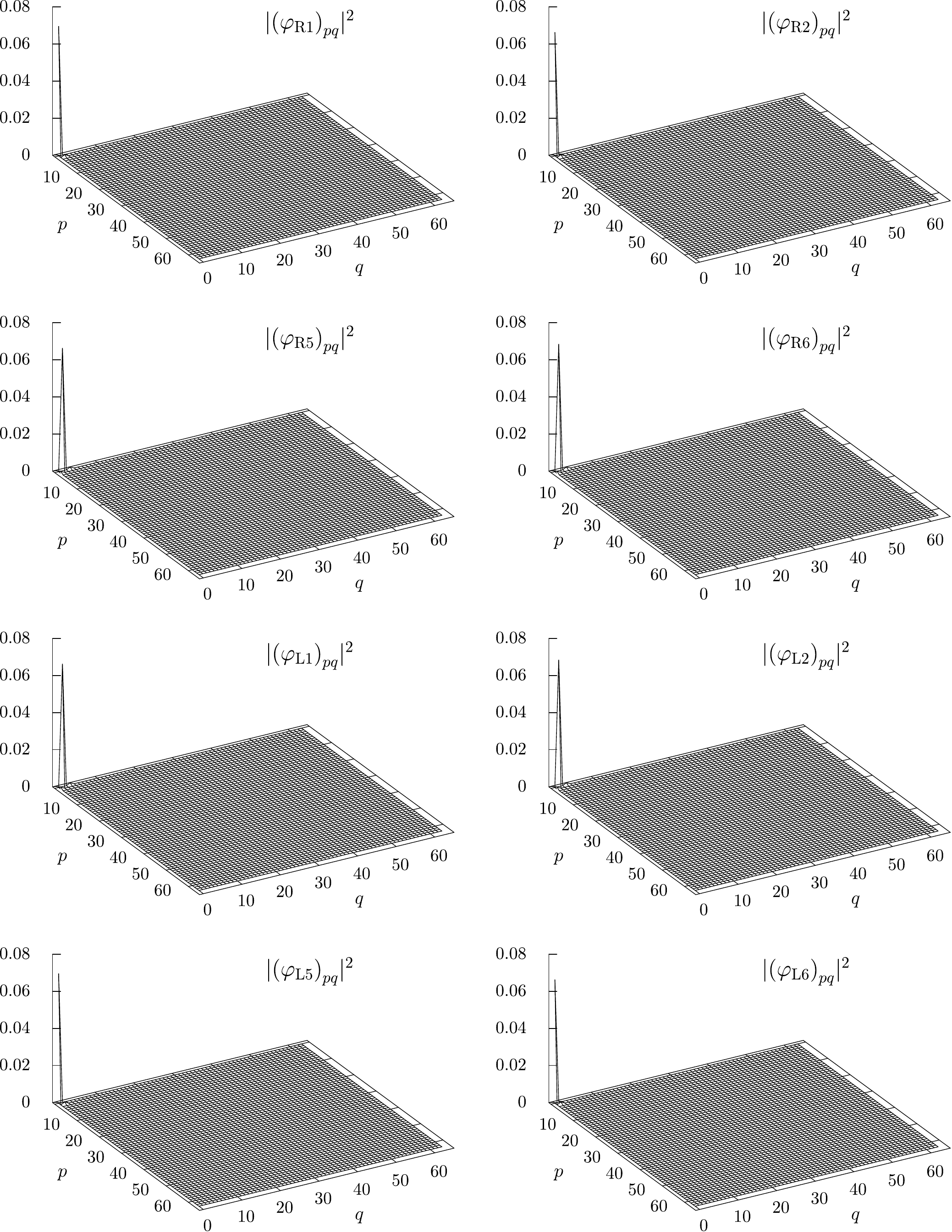}
\caption{The spinor components $\alpha = 1,2,5,6$ 
of the wave functions
$|(\varphi_{\mathrm{R}\alpha})_{pq}|^2$ 
and $|(\varphi_{\mathrm{L}\alpha})_{pq}|^2$
$(p=1,\ldots,N_{\rm Y}^{(1)},\ q=1,\ldots,N_{\rm Y}^{(2)})$ 
with the ``2d--3d'' ansatz are plotted for $N_{\rm Y}^{(1)}=N_{\rm Y}^{(2)}=64$. 
}
\label{wavefn_2d-3d}
\end{figure}

Next we consider the wave function corresponding 
to one of the 2 lowest 
eigenvalues for one of the 16 cases with $N_{\rm Y}^{(1)}=N_{\rm Y}^{(2)}=64$.
We calculate $|(\varphi_{\mathrm{R}\alpha})_{pq}|^2$ 
for each $\alpha$
and find that the wave function almost vanishes
except for the spinor components $\alpha=1,2,5,6$.
Similarly, we calculate $|(\varphi_{\mathrm{L}\alpha})_{pq}|^2$ 
for each $\alpha$,
and find that the wave function almost vanishes
except for the spinor components $\alpha=1,2,5,6$.
In Fig.~\ref{wavefn_2d-3d},
we plot $|(\varphi_{\mathrm{R}\alpha})_{pq}|^2$
and  $|(\varphi_{\mathrm{L}\alpha})_{pq}|^2$
for $\alpha=1,2,5,6$, the components of the 
wave functions that have nonzero elements. 
We find that 
$\varphi_{\mathrm{R}\alpha}$ is
localized at the $(1,1)$ element for $\alpha=1,2$ and
at the $(1,2)$ element for $\alpha=5,6$,
while 
$\varphi_{\mathrm{L}\alpha}$ is
localized at the $(1,2)$ element for $\alpha=1,2$ and
at the $(1,1)$ element for $\alpha=5,6$.
These results are consistent with the picture that 
the 2d brane and the 3d brane do not intersect, but
the points on each brane separated by the minimum distance
are uniquely determined.



\subsection{The cases with intersection but not at a point}

\paragraph{``3d--4d'' ansatz}
\label{subsec: 3d-4d}
For the first brane $Y_a^{(1)}$,
we set $Y_4^{(1)} = Y_5^{(1)} = Y_6^{(1)} = 0$
and use 4 3d solutions,
while for the second brane $Y_a^{(2)}$,
we set $Y_1^{(2)} = Y_2^{(2)} = 0$
and use 4 4d solutions.
Thus we consider $4\times 4=16$ backgrounds
with the matrix size
$N_{\rm Y}^{(1)}=N_{\rm Y}^{(2)}=32$, $48$ and $64$,
on which we solve the eigenvalue problem (\ref{eq: Diraceq_6d}).

\begin{figure}
\centering
\includegraphics[width=10.0cm]{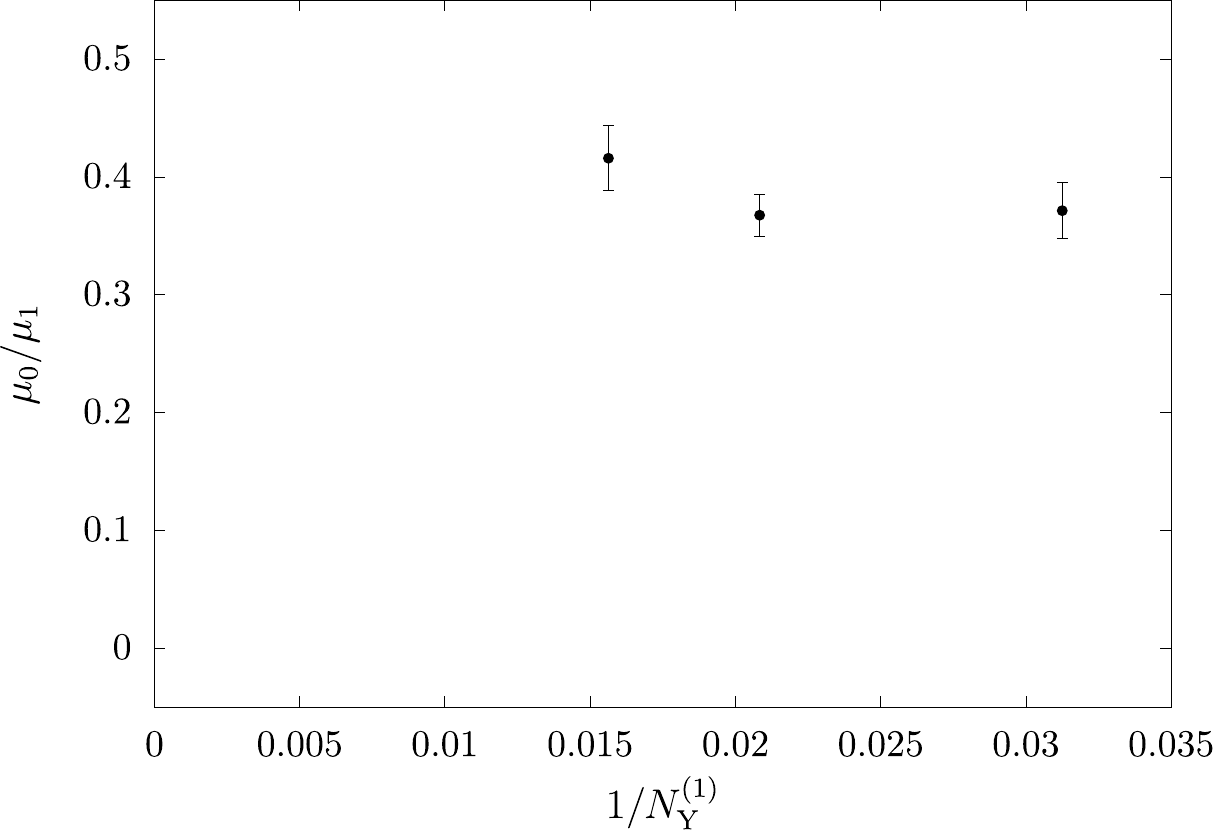}
\caption{The ratio $\mu_0/\mu_1$
with the ``3d--4d'' ansatz is plotted against $1/N_{\rm Y}^{(1)}$
for $N_{\rm Y}^{(1)}=N_{\rm Y}^{(2)}=32$, $48$ and $64$.}
\label{ratio_EV_3d-4d_NY1=[32-48-64]}
\end{figure}

In Fig.~\ref{ratio_EV_3d-4d_NY1=[32-48-64]},
we plot the ratio $\mu_0/\mu_1$
against $1/N_{\rm Y}^{(1)}$ for $N_{\rm Y}^{(1)}=N_{\rm Y}^{(2)}=32$, $48$ and $64$.
They do not converge to 0 in the $N_{\rm Y}^{(1)} \to \infty$ limit,
which implies that we do not obtain zero modes in this case.

\begin{figure}
\centering
\includegraphics[width=16.5cm]{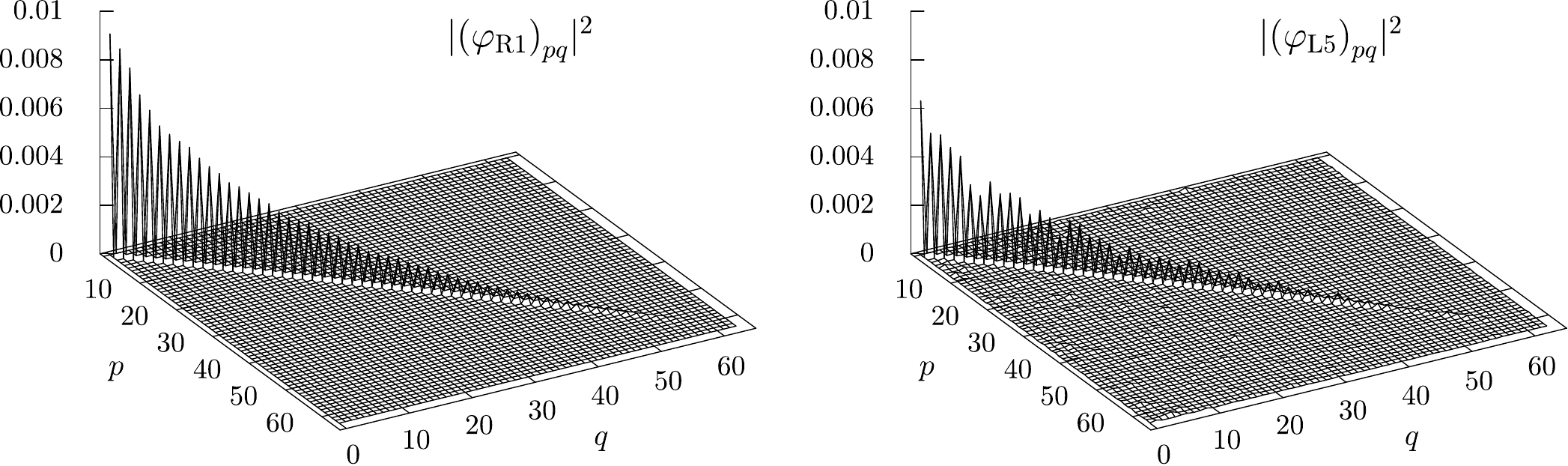}
\caption{The wave functions
$|(\varphi_{\mathrm{R}1})_{pq}|^2$ 
and $|(\varphi_{\mathrm{L}5})_{pq}|^2$ 
$(p=1,\ldots,N_{\rm Y}^{(1)},\ q=1,\ldots,N_{\rm Y}^{(2)})$ 
with the ``3d--4d'' ansatz are plotted
for $N_{\rm Y}^{(1)}=N_{\rm Y}^{(2)}=64$.  
}
\label{wavefn_3d-4d}
\end{figure}

Next we consider the wave function corresponding 
to the lowest eigenvalue for one of the 16 cases with $N_{\rm Y}^{(1)}=64$.
We calculate $|(\varphi_{\mathrm{R}\alpha})_{pq}|^2$ 
for each $\alpha$
and find that the wave function almost vanishes
except for the spinor component $\alpha=1$.
Similarly, we calculate $|(\varphi_{\mathrm{L}\alpha})_{pq}|^2$ 
for each $\alpha$,
and find that the wave function almost vanishes
except for the spinor component $\alpha=5$.
In Fig.~\ref{wavefn_3d-4d},
we plot $|(\varphi_{\mathrm{R}1})_{pq}|^2$
and  $|(\varphi_{\mathrm{L}5})_{pq}|^2$, the component of the 
wave functions that have nonzero elements. 
We find that the wave functions are not localized but have a long
tail along the diagonal line.
These results are consistent with the picture that 
the intersection does not occur at a point.
The zero modes do not appear, and the 
wave functions corresponding to the lowest eigenvalue do not localize.
Similar behaviors are observed for the ``4d--4d'' ansatz.

\section{Summary and discussions}
\label{conlusion}
\setcounter{equation}{0}
In this paper, we have proposed a numerical method
which enables us to solve the classical equation of motion of 
the type IIB matrix model.
In particular, we impose
a quasi-direct-product structure \eqref{eq: ansatz of A},
which is the most general ansatz compatible with the 
SO(3,1) Lorentz invariance.
Based on the solutions obtained in this way,
we have investigated the space--time structure in 
the (3+1) dimensions and 
the structure in the extra 6 dimensions.

First, we have focused on the (3+1)-dimensional space--time structure, 
which is represented by $X_\mu$ in Eq.~\eqref{eq: ansatz of A}.
When $X_0$ is diagonalized, $X_i$ become band-diagonal, 
which enables us to extract the time evolution.
We find that the 3-dimensional space expands with time,
and moreover it is smooth
unlike the singular structure observed
by simulations with an approximation to avoid the sign 
problem \cite{Aoki:2019tby}.
Our results support the conjecture in Ref.~\cite{Nishimura:2019qal}
that the singular structure is caused by the approximation
and that a smooth space--time should emerge in the $N \to \infty$ limit
if the sign problem is treated properly.

Next, we have focused on the structure in the extra dimensions.
In fact,
the existence of the matrix $M \neq {\bf 1}$
in the quasi-direct-product structure \eqref{eq: ansatz of A}
realizes naturally the appearance of 
intersecting D-branes represented by
a block-diagonal structure in the extra dimensions.
In order to confirm this picture, 
we have investigated
the eigenvalue spectrum of the Dirac operator
for two intersecting D-branes with various dimensionality.
In the ``3d--3d'' and ``2d--4d'' cases, 
our
solutions give rise to
Dirac zero modes in the limit of infinite matrix size.
We have also found that the wave functions corresponding 
to the lowest eigenvalue are localized at a point, 
which is consistent with the picture that the zero modes
appear from the intersection point.
It should be emphasized that
the zero modes were obtained 
for classical solutions of the type IIB matrix model
unlike the
previous 
studies \cite{Aoki:2010gv, Chatzistavrakidis:2011gs, Nishimura:2013moa, Aoki:2014cya},
where the configurations that give rise to zero modes are 
constructed by hand.
In the other cases in which the intersection either does not occur
or occurs but not at a point,
we do not obtain zero modes.

We consider that it is worth while to
extend the present studies
to larger matrix size.
By obtaining $X_\mu$ with larger $N_{\rm X}$, we can
determine how the space expands with time for a longer time period.
In particular, it would be interesting to see whether the expansion
is exponential or obeys a power law possibly depending on time.
It is also important to investigate whether the 3-dimensional space 
becomes uniform and SO(3) symmetric.
By obtaining $Y_a$ with larger $N_{\rm Y}$, we should be able
to see the emergence of zero modes more explicitly.
It would be also interesting to identify the zero modes
in the fluctuation of the matrices $Y_a$ around the classical solution,
which would correspond to the Higgs particles.
Then we can calculate the Yukawa couplings from the overlap of 
the wave functions 
of the Dirac zero modes and the Higgs particles.
Based on the information obtained in this way at the Planck scale,
we may use the renormalization group to see 
whether the Standard Model appears at low energy.
One of the most important issues to be addressed is whether
one can obtain chiral fermions.
We hope to report on the progress in these directions in the near future.

\section*{Acknowledgements}
The authors would like to thank H.\ Kawai and H.\ Steinacker 
for valuable discussions.
Computation was carried out on PC clusters 
at KEK and XC40 at YITP in Kyoto University.
This work was supported by computational time granted from the Greek
Research \& Technology Network (GRNET) in the 
National HPC facility -ARIS- under project ID LIKKT.
A.T.\ was supported in part by Grant-in-Aid for Scientific Research
(No. 18K03614) from Japan Society for the Promotion of Science.
K.H.\ is supported by Grant-in-Aid for JSPS Fellows (No.\ 19J10002).
\bibliography{ref_spacetime_chiralzeromode_classicalsolution_IIBMM}

\end{document}